\documentclass[english,preprint]{aastex}
\usepackage[T1]{fontenc}
\usepackage[latin9]{inputenc}
\usepackage{amsbsy}



\usepackage{babel}

\begin{document}

\title{Dispersion of Magnetic Fields in Molecular Clouds. I}

\author{Roger H. Hildebrand$^{1,2}$, Larry Kirby$^{1}$, Jessie L. Dotson$^{3}$,
Martin Houde$^{4}$, and John E. Vaillancourt$^{5}$}

\affil{$^{1}$Department of Astronomy and Astrophysics and Enrico Fermi
Institute, The University of Chicago, Chicago, IL 60637}

\affil{$^{2}$Department of Physics, The University of Chicago, Chicago,
IL 60637}

\affil{$^{3}$NASA Ames Research Center, Moffett Field, CA 94035}

\affil{$^{4}$Department of Physics and Astronomy, The University of Western
Ontario, London, ON, N6A 3K7, Canada}

\affil{$^{5}$Division of Physics, Mathematics, \& Astronomy, California
Institute of Technology, Pasadena, CA 91125}
\begin{abstract}
We describe a method for determining the dispersion of magnetic field
vectors about large-scale fields in turbulent molecular clouds. The
method is designed to avoid inaccurate estimates of magnetohydrodynamic
or turbulent dispersion - and help avoiding inaccurate estimates of
field strengths - due to large-scale, non-turbulent field structure
when using the well-known method of Chandrasekhar and Fermi. Our method
also provides accurate, independent estimates of the turbulent to
large-scale magnetic field strength ratio. We discuss applications
to the molecular clouds OMC-1, M17, and DR21(Main). 
\end{abstract}


\section{Introduction}

\citet{CF1953} used the dispersion of starlight polarization vectors
about contours of Galactic latitude \citep{Hiltner1949} $-$ together
with estimates of gas density and line-of-sight velocity dispersion
$-$ to determine the strength of the magnetic field in the arms of
the Galaxy. The same technique, {}``The Chandrasekhar-Fermi, `CF',
method'', has been applied, with modifications, to estimates of field
strengths in the relatively dense medium of molecular clouds at varying
temperature, wavelengths, sensitivities, and resolutions (e.g., \citealt{Lai2001, Lai2002, Lai2003, Crutcher2004, Houde2004, Girart2006, Curran2007}).

The basis for deriving field strengths from dispersion measurements
is the same for observations of Galactic arms or molecular clouds:
in either case dispersion decreases as the field strengthens. But
in the case of the Galactic arms, the dispersion is due to magnetohydrodynamic
(MHD) waves; the displacements are perpendicular to the direction
of propagation. In the case of turbulent dispersion in molecular clouds,
there is no preferred direction. The turbulent component can be in
any orientation.

Moreover, in dense clouds, the field may have structure due to effects
such as differential rotation, gravitational collapse, or expanding
\ion{H}{2} regions; i.e., structure not accounted for by the basic
CF analysis. Consequently, dispersion measured about mean fields,
assumed straight, may be much larger than should be attributed to
MHD waves or turbulence. Dispersion measured about model large-scale
fields \citep{Schleuning1998, Lai2002, Girart2006} that give approximate
fits to a polarization map will result in better estimates but still
give inaccurate values of the turbulent component, since they are
unlikely to perfectly match the true morphology of the large-scale
magnetic field. In this paper we describe a method for determining
magnetic field dispersion about local structured fields, without assuming
any model for the large-scale field. This method also provides accurate,
independent estimates of the turbulent to large-scale magnetic field
strength ratio.

We begin (\S\ref{sec:Difficulties}) by discussing difficulties one
must overcome in order to infer turbulent structure from polarization
maps, regardless of large-scale effects. In \S\ref{sec:Function},
we present the method and in \S\ref{sec:Applications}, we give applications
to the molecular clouds OMC-1, M17, and DR21(Main). Detailed derivations
resulting in the relations and functions used in the aforementioned
sections, as well as the data analysis, will be found in the appendices
at the end of the paper.

\section{Difficulties in Deriving the Turbulent Structure from Polarized Emission\label{sec:Difficulties}}

Turbulent velocities of gas motion within and between clumps of material
along the line of sight can often be inferred from the widths and
centers of molecular lines (e.g., \citealt{Kleiner1984, Kleiner1985, Kleiner1987}).
But dust polarization measurements of dispersion in magnetic field
direction do not separate contributions from either volume elements
located along the line of sight or across the area subtended by the
telescope beam. Hence the measured angular dispersion tends to be
a smoothed version of the true dispersion \citep{Myers1991, Wiebe2004}.
Nonetheless, a corresponding average of the dispersion remains and
is measurable; for a given object observations will thus reveal a
higher degree of dispersions when they are realized at an accordingly
higher spatial resolution. 

A potentially fruitful line of attack for estimating magnetic field
strengths relies on comparisons of observed and simulated maps of
the net polarization (e.g., \citealt{Ostriker2001, Heitsch2001, Falceta2008}).
If the simulations are computed for the resolution, column density,
and other characteristics of the cloud under study, and if they are
computed for several models of the key variables (e.g., field strength
and turbulent fraction), then one can find the model giving the best
fit to the observations. A valid simulation must also take into account
temperatures \citep{Vaillancourt2002} and grain alignment efficiencies
in different environments \citep{Hoang2008}. The comparisons are
facilitated if both the observations and the simulations are presented
in tables of Stokes parameters, so that each can be analyzed in the
same way. The various modifications of the CF method that have been
used to relate net dispersion to field strength (e.g., \citealt{Ostriker2001, Padoan2001, Heitsch2001, Kudoh2003, Houde2004})
are, in effect, first-order substitutes for simulations. 

But a meaningful comparison between simulations and observations can
only be achieved if a reliable estimate of the spatially averaged
angular dispersion can be secured experimentally. It would therefore
be advantageous if a more general method, which does not depend on
any assumption concerning the morphology of the large-scale magnetic
field, were devised. The method we describe in the following section
allows for the evaluation of the plane-of-the-sky turbulent angular
dispersion in molecular clouds while avoiding inaccurate estimates
of the turbulence and corresponding inaccurate estimates of field
strengths due to distortions in polarization position angles by large-scale
non-turbulent effects. This method can lead to valid estimates of
magnetic field strengths only under conditions such that the Chandrasekhar-Fermi
method can be properly applied: a smooth, low noise, polarization
map, precise measured densities and gas velocities that are moderately
uniform, and an adequate accounting of the integration process implicit
to polarization measurements. This latter aspect will be addressed
in a subsequent paper.

\section{A Function to Describe Dispersion about Large-scale Fields\label{sec:Function}}

Consider a map precisely showing the angle $\Phi\mathbf{\left(\mathbf{x}\right)}$
of the (two-dimensional) plane-of-the-sky projected magnetic field
vector $\mathbf{B\left(\mathbf{x}\right)}$ at many points in a molecular
cloud. We obtain a measure of the difference in angle, $\Delta\Phi\left(\ell\right)\equiv\Phi\left(\mathbf{x}\right)-\Phi\left(\mathbf{x}+\boldsymbol{\ell}\right)$,
between the $N(\mathbf{\ell)}$ pairs of vectors separated by displacements
$\boldsymbol{\ell}$, also restricted to the plane-of-the-sky, through
the following function

\begin{equation}
\left\langle \Delta\Phi^{2}\left(\ell\right)\right\rangle ^{1/2}\equiv\left\{ \frac{1}{N\left(\ell\right)}\sum_{i=1}^{N\left(\ell\right)}\left[\Phi\left(\mathbf{x}\right)-\Phi\left(\mathbf{x}+\boldsymbol{\ell}\right)\right]^{2}\right\} ^{1/2},\label{eq:func}\end{equation}

\noindent where $\left\langle \cdots\right\rangle $ denotes an average
and $\ell=\left|\boldsymbol{\ell}\right|$. The square of equation
(\ref{eq:func}) is also often referred to as a {}``structure function''
(of the second order in this case; see \citealt{Falceta2008, Frisch1995}),
but for our applications we shall refer to it as the {}``dispersion
function'' and assume that it is isotropic (i.e., it only depends
on the magnitude of the displacement, $\boldsymbol{\ell}$, and not
its orientation). We seek to determine how this quantity varies as
a function of $\ell$. 

To do so, we will assume that the magnetic field $\mathbf{B\left(\mathbf{x}\right)}$
is composed of a large-scale, structured field, $\mathbf{B_{0}\left(\mathbf{x}\right)}$,
and a turbulent (or random) component, $\mathbf{B_{\mathrm{t}}\left(\mathbf{x}\right)}$,
which are statistically independent. We also limit ourselves to cases
where $\delta<\ell\ll d$, where $\delta$ is the correlation length
characterizing $\mathbf{B_{\mathrm{t}}\left(\mathbf{x}\right)}$ and
$d$ is the typical length scale for variations in $\mathbf{B_{0}\left(\mathbf{x}\right)}$.

Focusing on $\mathbf{B_{0}\left(\mathbf{x}\right)}$ we would expect
its contribution to the dispersion function to increase (since $\left\langle \Delta\Phi^{2}\left(\ell\right)\right\rangle $
is positive definite) almost linearly starting at $\ell=0$ and for
small displacements $\ell\ll d$, as would be expected from the Taylor
expansion of any smoothly varying quantity. We denote by $m$ the
slope characterizing this linear behavior. We also expect a contribution
from the turbulent component of the magnetic field $\mathbf{B_{\mathrm{t}}\left(\mathbf{x}\right)}$.
This contribution will vary from zero as $\ell\rightarrow0$ (when
the two magnetic field vectors are co-aligned) to a maximum average
value when the displacement exceeds the correlation length $\delta$
characterizing $\mathbf{B_{\mathrm{t}}\left(\mathbf{x}\right)}$.
More precisely, we expect that the turbulent contribution to the angular
dispersion will be a constant, which we denote by $b$, as long as
$\ell>\delta$. These two contributions must be combined quadratically,
since the large-scale and turbulent fields are statistically independent,
to yield

\begin{equation}
\left\langle \Delta\Phi^{2}\left(\ell\right)\right\rangle \simeq b^{2}+m^{2}\ell^{2}\label{eq:disp}\end{equation}

\noindent when $\delta<\ell\ll d$. 

A more formal and rigorous derivation of equation (\ref{eq:disp})
is established in Appendix \ref{sec:Dispersion} under the further
assumptions of homogeneity and isotropy in the magnetic field strength
over space. Although these assumptions are unlikely to be realized
across molecular clouds, this level of idealization is necessary to
allow us to gain insights on, and some quantitative measure of, the
importance of the turbulent component of the magnetic field in molecular
clouds.

In reality, the measured dispersion function from a polarization map
will also include a contribution, $\sigma_{\mathrm{M}}\left(\ell\right)$,
due to measurement uncertainties on the polarization angles $\Phi\left(\mathbf{x}\right)$
that must be added (quadratically) to equation (\ref{eq:disp}). The
square of the total measured dispersion function then becomes

\begin{equation}
\left\langle \Delta\Phi^{2}\left(\ell\right)\right\rangle _{\mathrm{tot}}\simeq b^{2}+m^{2}\ell^{2}+\sigma_{\mathrm{M}}^{2}\left(\ell\right).\label{eq:disp_m}\end{equation}

\noindent when $\delta<\ell\ll d$. The function $\left\langle \Delta\Phi^{2}\left(\ell\right)\right\rangle _{\mathrm{tot}}$,
not $\left\langle \Delta\Phi^{2}\left(\ell\right)\right\rangle $,
is the one calculated from a polarization map (from an averaging process
similar to equation {[}\ref{eq:func}{]}), and will thus contain separate
components due to the large-scale structure (i.e., $m\ell$), the
turbulent dispersion about the large-scale field (i.e., $b$, the
quantity we wish to measure), and measurement uncertainties (i.e.,
$\sigma_{\mathrm{M}}\left(\ell\right)$).

\begin{figure}
\plotone{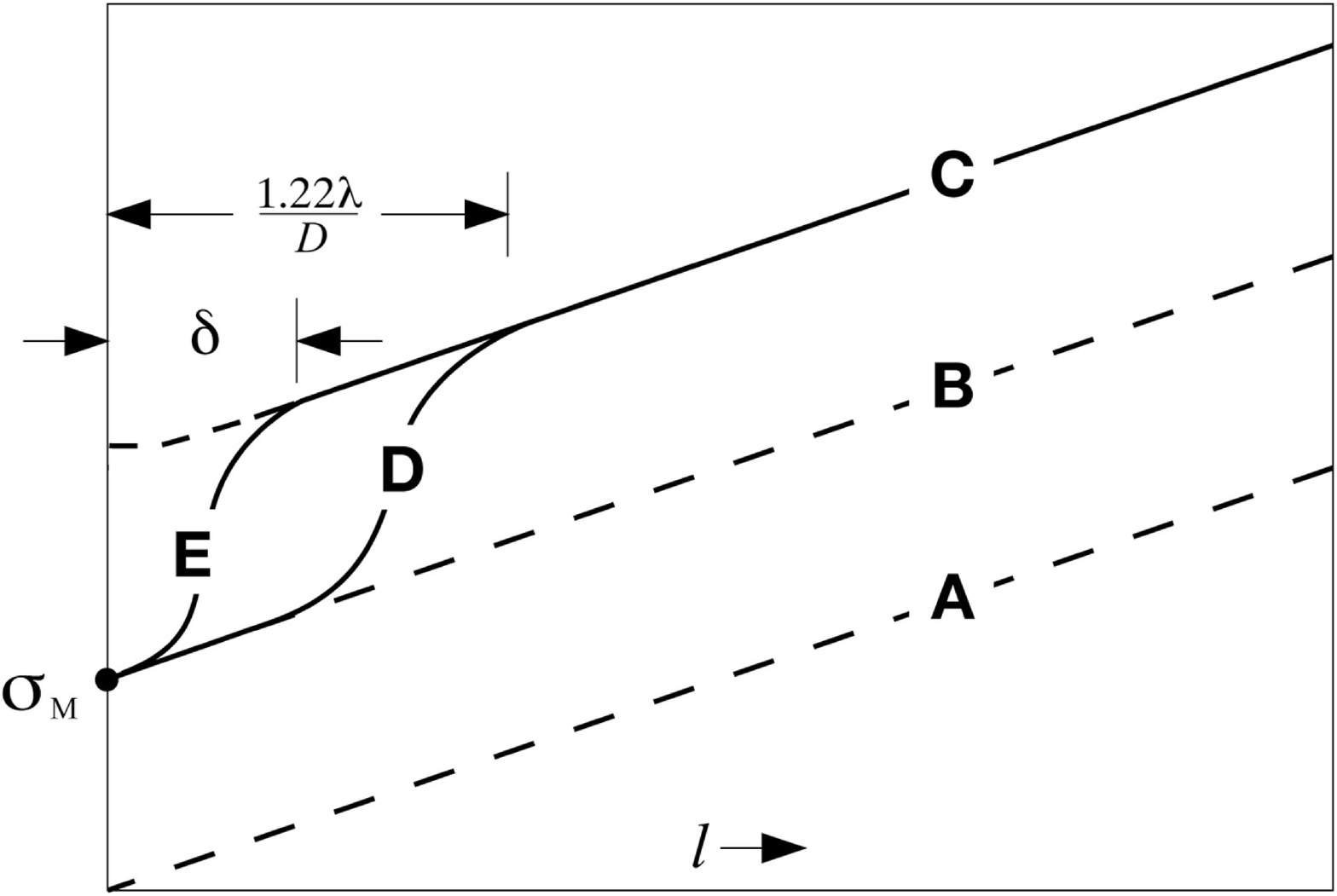}

\caption{\label{fig:ideal}Dispersion: Idealized plots of the angular dispersion
function, $\left\langle \Delta\Phi^{2}\left(\ell\right)\right\rangle ^{1/2}$,
between pairs of magnetic field vectors separated by displacements
$\mathbf{\ell}$, for values of $\mathbf{\ell}\ll d$, with $d$ the
typical length scale for variations in the large-scale magnetic field
(see \S\ref{sec:Function}). Curve A: no measurement uncertainty;
no turbulence. Curve B: with measurement uncertainty, $\sigma_{\mathrm{M}}$.
Curve C: with turbulence. Curves D and E: accounting for correlation
in polarization angles at displacements $\mathbf{\ell}$ smaller than
the larger of the telescope beam ($1.22\lambda/D$) (Curve D) or the
turbulent correlation length $\delta$ (Curve E). }

\end{figure}

If there were no turbulence and no measurement uncertainties, then,
for $\mathbf{\ell}\ll d$ the measured dispersion function would be
a straight line with zero intercept, $\left\langle \Delta\Phi\left(\ell\right)^{2}\right\rangle _{\mathrm{tot}}^{1/2}=m\mathbf{\ell}$
(see Figure \ref{fig:ideal}, Curve A). Taking the measurement uncertainty,
$\sigma_{\mathrm{M}}\left(\ell\right)$, into account, the line would
be displaced upward as specified by equation (\ref{eq:disp_m}) (Curve
B, where $\sigma_{\mathrm{M}}$ was assumed to be independent of $\ell$).
Likewise when we next consider turbulence, the curve will again be
displaced upward in the same manner (Curve C) \textit{\emph{except}}
at values of $\mathbf{\ell}$ below the angular resolution scale at
which the observations were made (Curve D), or below the turbulent
correlation scale $\delta$ (Curve E). Theoretical and observational
estimates of $\delta$ for molecular clouds are on the order of 1
mpc (\citealt{Lazarian2004, Li2008b}, respectively), well below the
size of the telescope beam with which the observations presented in
this paper were obtained. Although it has not yet been feasible to
resolve $\delta$, it \textit{\emph{is}} now feasible to determine
the turbulent dispersion at scales comparable to the approximately
linear portion of $\left\langle \Delta\Phi\left(\ell\right)^{2}\right\rangle _{\mathrm{tot}}^{1/2}$.

Notice that $\sigma_{\mathrm{M}}\left(\ell\right)$ can be accurately
determined through the uncertainties on the measured polarization
angles of each pair of points used in the calculation of $\left\langle \Delta\Phi\left(\ell\right)^{2}\right\rangle _{\mathrm{tot}}$,
and by then subtracting its square to obtain $\left\langle \Delta\Phi\left(\ell\right)^{2}\right\rangle $.
As the number and precision of the vectors improve, equation (\ref{eq:disp})
can be fitted to the data for $\delta<\ell\ll d$, and the intercept
at $\ell=0$ provides us with the turbulent contribution, $b^{2}$,
to the square of the angular dispersion. 

The Chandrasekhar-Fermi method for evaluating strength of the plane-of-the-sky
component of the large-scale magnetic field \citep{CF1953} implies
that

\begin{equation}
\frac{\delta B}{B_{0}}\simeq\frac{\sigma\left(v\right)}{V_{\mathrm{A}}},\label{eq:ratio_CF}\end{equation}

\noindent where $\delta B$ stands for the variation in the magnetic
field about the large-scale field $B_{0}$, $\sigma\left(v\right)$
is the one-dimensional velocity dispersion of the gas (of mass density
$\rho$) coupled to the magnetic field, and 

\begin{equation}
V_{\mathrm{A}}=\frac{B_{0}}{\sqrt{4\pi\rho}}\label{eq:Alfven}\end{equation}

\noindent is the Alfvén speed. It is further assumed that the dispersion,
$\sigma_{\Phi}$, in the polarization angles $\Phi\left(\mathbf{x}\right)$
across a map is given by 

\begin{equation}
\sigma_{\Phi}\simeq\frac{\delta B}{B_{0}}.\label{eq:sig_phi}\end{equation}

\noindent The combination of equations (\ref{eq:ratio_CF}), (\ref{eq:Alfven}),
and (\ref{eq:sig_phi}) allows for the aforementioned determination
of the plane-of-the-sky component of the large-scale magnetic field
strength as a function of $\rho$, $\sigma\left(v\right)$ (determined
from the width of appropriate spectral line profiles), and $\sigma_{\Phi}$
(determined from polarization measurements). 

It is shown with equation (\ref{eq:ratio}) in Appendix \ref{sec:Dispersion}
that the ratio of the turbulent to large-scale magnetic field strength
is given by

\begin{equation}
\frac{\left\langle B_{\mathrm{t}}^{2}\right\rangle ^{1/2}}{B_{0}}=\frac{b}{\sqrt{2-b^{2}}}.\label{eq:ratio_t2m}\end{equation}

\noindent It is therefore apparent that we should make the correspondence
$\left\langle B_{\mathrm{t}}^{2}\right\rangle ^{1/2}\rightarrow\delta B$
and that

\begin{eqnarray}
B_{0} & \simeq & \sqrt{\left(2-b^{2}\right)4\pi\rho}\,\frac{\sigma\left(v\right)}{b}\nonumber \\
 & \simeq & \sqrt{8\pi\rho}\,\frac{\sigma\left(v\right)}{b},\label{eq:B0_app}\end{eqnarray}

\noindent where the last equation applies when $B_{\mathrm{t}}\ll B_{0}$.
The fact that the turbulent dispersion, $b$, is to be divided by
approximately $\sqrt{2}$ before being inserted the Chandrasekhar-Fermi
equation is readily understood by the fact that (neglecting the contribution
of the large-scale field)

\begin{eqnarray*}
\left\langle \Delta\Phi^{2}\left(\ell\right)\right\rangle  & = & \left\langle \left[\Phi\left(\mathbf{x}\right)-\Phi\left(\mathbf{x}+\boldsymbol{\ell}\right)\right]^{2}\right\rangle \\
 & = & 2\left(\left\langle \Phi^{2}\right\rangle -\left\langle \Phi\right\rangle ^{2}\right)\\
 & = & 2\sigma_{\Phi}^{2},\end{eqnarray*}

\noindent when $\ell>\delta$. Since we also know that $\left\langle \Delta\Phi^{2}\left(\ell\right)\right\rangle =b^{2}$
at these scales, we then find that $b^{2}=2\sigma_{\Phi}^{2}$, which
is consistent with equations (\ref{eq:sig_phi}) and (\ref{eq:ratio_t2m}).

It should be noted that the combination of equations (\ref{eq:ratio_t2m})
and (\ref{eq:B0_app}) allows, in principle, for the determination
of both the large-scale and turbulent magnetic fields' strength from
polarization and spectroscopy data.

\section{Applications to the Molecular Clouds OMC-1, M17, and DR21(Main)\label{sec:Applications}}

Using data from the polarimeter Hertz (Dowell et al. 1998) at the
Caltech Submillimeter Observatory at 350 $\mu\mathrm{m}$, we have
measured dispersion functions for the molecular clouds OMC-1, M17,
and DR21(Main). These data are discussed in detail in \citet{Houde2004b}
for OMC-1, \citet{Houde2002} for M17, and \citet{Kirby2009} for
DR21(Main). Figure \ref{fig:data} shows the results for all sources.
More details on the data analysis will be found in Appendix \ref{sec:Data}.

For each object, we show $\left\langle \Delta\Phi^{2}\left(\ell\right)\right\rangle ^{1/2}$
over the cloud along with the best fit from equation (\ref{eq:disp})
using the first three data points to ensure that $\ell\ll d$, as
much as possible. The measurement uncertainties were removed prior
to operating the fits to the corresponding data sets. The turbulent
contribution to the total angular dispersion is determined by the
zero intercept of the fit to the data at $\ell=0$. The net turbulent
component, $b$, is $0.18\pm0.01$ rad ($10.4\pm0.6\,\deg$), $0.12\pm0.02$
rad ($6.8\pm1.3\,\deg$), and $0.15\pm0.01$ rad ($8.3\pm0.3\,\deg$)
for M17, DR21(Main), and OMC-1, respectively. 

Although large variations in density within the observed regions prevent
a reliable estimate in the field strength at precise locations, it
is still possible to give some average value for the large-scale and
turbulent field strengths. To do so we use representative line width
measurements from H$^{13}$CO$^{+}$ $J=3\rightarrow2$ detections
within the three clouds. For OMC-1 and M17 we have used the corresponding
measurements published in \citet{Houde2000} (more precisely, an average
of the variances obtained at the two positions listed for M17), while
for DR21(Main) we have used previously unpublished data. This molecular
species is well suited for this as the effective density needed for
line detection with the aforementioned transition ($n_{\mathrm{eff}}\sim10^{5}\,\mathrm{cm}^{-3}$,
see \citealt{Evans1999}) is close to the densities at which dust
continuum emission is detected at the measured wavelength. Also, the
corresponding spectral lines are likely to be optically thin (like
the dust continuum) and an ion molecule such as this one is better
coupled to the magnetic field (and the dust) than corresponding neutral
species (e.g., H$^{13}$CN for the same rotational transition) over
the whole turbulent energy density spectrum \citep{Li2008b}. Therefore,
using a density of $10^{5}\,\mathrm{cm}^{-3}$ and a mean molecular
weight of 2.3 we obtain the results shown in Table \ref{tab:Results}.
As a simple comparison, the values of dispersion shown in the table
are approximately three times lower than would be obtained if one
naively calculated the dispersions about the global mean field (i.e.,
the field direction defined by the mean of all polarization vectors
in the corresponding map). More precisely, we get dispersions of 27.2,
21.0, and 26.8 degrees about the global mean field orientation for
M17, DR21(Main), and OMC-1, respectively.

We wish to emphasize the fact that the quoted values for $B_{0}$
could not be precise to better than a factor of a few due to a lack
of precise gas density numbers. Moreover,  the values for the large-scale
magnetic field strength we derived are up to an order of magnitude
higher than those obtained with other observational means (cf., the
results of \citet{Crutcher1999} for OMC-1 and M17 using CN Zeeman
measurements). These high values are in part the result of the smaller
angular dispersions obtained using our technique as compared to more
common methods used when applying the Chandrasekhar-Fermi equation
(e.g., model fits to large-scale fields). One must keep in mind, however,
that the process of signal integration through the thickness of the
cloud and across the telescope beam that is inherent to polarization
measurements has not been taken into account. We will show in a subsequent
publication how this situation is rectified when these considerations
(and others) are carefully taken into account \citep{Myers1991, Ostriker2001, Wiebe2004}.
Nevertheless, the turbulent to large-scale magnetic field strength
ratio is precisely evaluated through our equation (\ref{eq:ratio_t2m}).

\begin{deluxetable}{lcccc}

\tablewidth{0pt}
\tablecolumns{5}

\tablehead{
\colhead{Object} & \colhead{$b$\tablenotemark{a}} & 
\colhead{$\left\langle B_{\mathrm{t}}^{2}\right\rangle ^{1/2}/B_{0}$\tablenotemark{b}} &
\colhead{$\sigma(v)$} & \colhead{$B_{0}$\tablenotemark{c}} \\
\colhead{} & \colhead{(deg)} & \colhead{} & \colhead{(km s$^{-1}$)} 
& \colhead{(mG)}
}

\tablecaption{Results for the dispersion, the turbulent to mean magnetic field strength ratio, the line widths, and the mean field strength.  \label{tab:Results}}

\startdata

OMC-1 & $8.3\pm0.3$ & $0.10\pm0.01$ & 1.85 & 3.8 \\
M17 & $10.4\pm0.6$ & $0.13\pm0.01$ & 1.66 & 2.9 \\ 
DR21(Main) & $6.8\pm1.3$ & $0.08\pm0.02$ & 4.09 & 10.6

\enddata

\tablenotetext{a}{Turbulent dispersion (i.e., the dispersion limit as $\ell\rightarrow0$).}
\tablenotetext{b}{Calculated with equation (\ref{eq:ratio_t2m}).}
\tablenotetext{c}{Calculated with equation (\ref{eq:B0_app}),  assumes a density of $10^5$ cm$^{-3}$ and a mean molecular weight of 2.3. These estimates are not precise to better than a factor of a few. The process of signal integration through the thickness of the cloud and across the telescope beam inherent to the polarization measurements has also not been taken into account.}

\end{deluxetable}

\section{Summary}

We have described a method to estimate plane-of-the-sky turbulent
dispersion in molecular clouds while avoiding inaccurate estimates
of the turbulence and corresponding inaccurate estimates of field
strengths due to distortions in polarization position angles by large-scale
non-turbulent effects. The method does not depend on any model of
the large-scale field. We plot a {}``dispersion function'', the
mean absolute difference in angle between pairs of vectors as a function
of their displacement $\mathbf{\ell}$ and show that this function
increases approximately linearly for displacements greater than the
instrument resolution, greater than the correlation length, $\delta$,
and less than the typical length scale, $d$, for variations in the
large-scale magnetic field (\S4). We emphasize that this method can
lead to valid estimates of magnetic field strengths only under conditions
such that the Chandrasekhar-Fermi method can be properly applied:
a smooth, low noise, polarization map, precise measured densities
and gas velocities that are moderately uniform, and an adequate accounting
of the integration process implicit to polarization measurements.
This method, however, provides accurate estimates of the turbulent
to large-scale magnetic field strength ratio.

Although the resolution of the instruments now available are not adequate
to directly determine the correlation length, $\delta$, one can still
determine the dispersion in the fields at scales where $\delta<\ell\ll d$
for the angular dispersion function. We have successfully done this
for the OMC-1, M17, and DR21(Main) molecular clouds.

\acknowledgements{We thank Shantanu Basu for helpful discussions. This work has been
supported in part by NSF grants AST 05-05230, AST 02-41356, and AST
05-05124. L.K. acknowledges support from the Department of Astronomy
and Astrophysics of the University of Chicago. M.H.'s research is
funded through the NSERC Discovery Grant, Canada Research Chair, Canada
Foundation for Innovation, Ontario Innovation Trust, and Western's
Academic Development Fund programs. J.E.V. acknowledges support from
the CSO, which is funded through NSF AST 05-40882. }

\appendix
\section{Dispersion Relation Derivation}\label{sec:Dispersion}

\subsection{Analysis in Three Dimensions}

Let us define the total magnetic field $\mathbf{B\left(\mathbf{x}\right)}$
as being composed of a deterministic, $\mathbf{B_{0}\left(\mathbf{x}\right)}$,
and a turbulent (or random), $\mathbf{B_{\mathrm{t}}\left(\mathbf{x}\right)}$,
components such that 

\begin{equation}
\mathbf{\mathbf{B\left(\mathbf{x}\right)=}B_{0}\mathbf{\mathbf{\left(\mathbf{x}\right)+\mathbf{B}_{\mathrm{t}}\left(\mathbf{x}\right)}}}.\label{eq:Btot}\end{equation}

\noindent These quantities have the following averages at points $\mathbf{x}$
and $\mathbf{y}$\begin{eqnarray}
\left\langle \mathbf{B}_{0}\left(\mathbf{x}\right)\right\rangle  & = & \mathbf{B}_{0}\left(\mathbf{x}\right)\nonumber \\
\left\langle \mathbf{B}_{0}\left(\mathbf{x}\right)\cdot\mathbf{B}_{\mathrm{0}}\left(\mathbf{y}\right)\right\rangle  & = & \mathbf{B}_{0}\left(\mathbf{x}\right)\cdot\mathbf{B}_{\mathrm{0}}\left(\mathbf{y}\right)\nonumber \\
\left\langle \mathbf{B}_{\mathrm{t}}\left(\mathbf{x}\right)\right\rangle  & = & 0\nonumber \\
\left\langle \mathbf{B}_{0}\left(\mathbf{x}\right)\cdot\mathbf{B}_{\mathrm{t}}\left(\mathbf{y}\right)\right\rangle  & = & \left\langle \mathbf{B}_{0}\left(\mathbf{x}\right)\right\rangle \cdot\left\langle \mathbf{B}_{\mathrm{t}}\left(\mathbf{y}\right)\right\rangle =0.\label{eq:averages}\end{eqnarray}

\noindent We will further assume homogeneity in the field strength
over space. That is,

\begin{eqnarray}
\left\langle \mathbf{B}_{0}^{2}\left(\mathbf{x}\right)\right\rangle  & = & \left\langle \mathbf{B}_{0}^{2}\left(\mathbf{y}\right)\right\rangle =B_{0}^{2}\nonumber \\
\left\langle \mathbf{B}_{\mathrm{t}}^{2}\left(\mathbf{x}\right)\right\rangle  & = & \left\langle \mathbf{B}_{\mathrm{t}}^{2}\left(\mathbf{y}\right)\right\rangle =\left\langle \mathbf{B}_{\mathrm{t}}^{2}\right\rangle .\label{eq:homo}\end{eqnarray}

Let us now consider the quantity 

\begin{equation}
\left\langle \cos\left[\Delta\Phi_{\mathrm{3D}}\left(\boldsymbol{\ell}\right)\right]\right\rangle \equiv\frac{\left\langle \mathbf{B}\left(\mathbf{x}\right)\cdot\mathbf{B}\left(\mathbf{x+\boldsymbol{\ell}}\right)\right\rangle }{\left[\left\langle B^{2}\left(\mathbf{x}\right)\right\rangle \left\langle B^{2}\left(\mathbf{x}+\boldsymbol{\ell}\right)\right\rangle \right]^{1/2}}.\label{eq:cos(dphi)}\end{equation}

\noindent The quantity $\Delta\Phi_{\mathrm{3D}}\left(\mathbf{\boldsymbol{\ell}}\right)$
is the angle difference between two magnetic field (or polarization)
vectors separated by a distance $\boldsymbol{\ell}$, the average
of its square is the function that we wish to evaluate through polarization
measurements (albeit in two dimensions, see \S\ref{sec:2D}). Using
equations (\ref{eq:Btot}) and (\ref{eq:averages}) we find that the
numerator of equation (\ref{eq:cos(dphi)}) (i.e., the autocorrelation
of the total magnetic field; see \citealt{Frisch1995}) becomes

\begin{equation}
\left\langle \mathbf{B}\left(\mathbf{x}\right)\cdot\mathbf{B}\left(\mathbf{x+\boldsymbol{\ell}}\right)\right\rangle =B_{0}^{2}+\left\langle \mathbf{B}_{0}\left(\mathbf{x}\right)\cdot\left[\sum_{n=1}^{\infty}\frac{\ell^{n}}{n!}\left(\mathbf{e}_{\ell}\cdot\nabla\right)^{n}\mathbf{B}_{0}\left(\mathbf{x}\right)\right]\right\rangle +\left\langle \mathbf{B}_{\mathrm{t}}\left(\mathbf{x}\right)\cdot\mathbf{B}_{\mathrm{t}}\left(\mathbf{x+\boldsymbol{\ell}}\right)\right\rangle ,\label{eq:num}\end{equation}

\noindent where we used the Taylor expansion

\begin{equation}
\mathbf{B}_{0}\left(\mathbf{x+\boldsymbol{\ell}}\right)=\mathbf{B}_{0}\left(\mathbf{x}\right)+\sum_{n=1}^{\infty}\frac{\ell^{n}}{n!}\left(\mathbf{e}_{\ell}\cdot\nabla\right)^{n}\mathbf{B}_{0}\left(\mathbf{x}\right),\label{eq:taylor}\end{equation}

\noindent with $\mathbf{e}_{\ell}$ the unit vector in the direction
of $\boldsymbol{\ell}$. 

If we introduce $d$ the scale length characterizing (large-scale)
variations in $\mathbf{B}_{0}$ and we consider situations where $\ell=\left|\boldsymbol{\ell}\right|\ll d$,
then we would expect that only the first term in the summation on
the right hand side of equation (\ref{eq:taylor}) would need to be
retained. If we define $\varphi_{i}$ as the angle between the gradient
of the $i$-component (i.e., $i=x,y,z$) of $\mathbf{B}_{0}$ and
$\mathbf{e}_{\ell}$, then when averaging over a large polarization
map we have

\begin{equation}
\left\langle B_{0,i}\left(\mathbf{x}\right)\left[\ell\left(\mathbf{e}_{\ell}\cdot\nabla\right)B_{0,i}\left(\mathbf{x}\right)\right]\right\rangle =\ell B_{0,i}\left(\mathbf{x}\right)\left|\nabla B_{0,i}\right|\left\langle \cos\left(\varphi_{i}\right)\right\rangle .\label{eq:1st_term}\end{equation}

\noindent But since $\mathbf{e}_{\ell}$ is equally likely to be oriented
in any direction over the whole map we have $\left\langle \cos\left(\varphi_{i}\right)\right\rangle =0$
and the first order term of the Taylor expansion (i.e., equation {[}\ref{eq:1st_term}{]})
cancels out. It therefore follows that the first non-vanishing term
in the summation on the right hand side of equation (\ref{eq:taylor})
is of second order with 

\begin{equation}
\left\langle \mathbf{B}_{0}\left(\mathbf{x}\right)\cdot\left[\sum_{n=1}^{\infty}\frac{1}{n!}\left(\boldsymbol{\ell}\cdot\nabla\right)^{n}\mathbf{B}_{0}\left(\mathbf{x}\right)\right]\right\rangle \simeq\frac{1}{2}\left\langle \mathbf{B}_{0}\left(\mathbf{x}\right)\cdot\left(\mathbf{e}_{\ell}\cdot\nabla\right)^{2}\mathbf{B}_{0}\left(\mathbf{x}\right)\right\rangle \ell^{2}.\label{eq:2nd_order}\end{equation}

\noindent when $\ell\ll d$. If we also assume stationarity for the
turbulent magnetic field, then we define the autocorrelation of the
turbulent field as 

\begin{equation}
\left\langle \mathbf{B}_{\mathrm{t}}\cdot\mathbf{B}_{\mathrm{t}}\left(\boldsymbol{\ell}\right)\right\rangle \equiv\left\langle \mathbf{B}_{\mathrm{t}}\left(\mathbf{x}\right)\cdot\mathbf{B}_{\mathrm{t}}\left(\mathbf{x+\boldsymbol{\ell}}\right)\right\rangle ,\label{eq:auto}\end{equation}

\noindent which, if we now define $\delta$ as the correlation length
for $\mathbf{B}_{\mathrm{t}}\left(\mathbf{x}\right)$, has the following
limits

\begin{equation}
\left\langle \mathbf{B}_{\mathrm{t}}\cdot\mathbf{B}_{\mathrm{t}}\left(\mathbf{\boldsymbol{\ell}}\right)\right\rangle =\left\{ \begin{array}{cl}
\left\langle B_{\mathrm{t}}^{2}\right\rangle , & \mathrm{when\,\,}\ell\rightarrow0\\
0, & \mathrm{when\,\,}\ell>\delta\end{array}\right.\label{eq:corr_Bt}\end{equation}

\noindent since the turbulent field is assumed uncorrelated over separations
exceeding $\delta$ and $\left\langle \mathbf{B}_{\mathrm{t}}\right\rangle =0$
from the third of equations (\ref{eq:averages}). Inserting equations
(\ref{eq:2nd_order}) and (\ref{eq:auto}) into equation (\ref{eq:num})
we have

\begin{equation}
\left\langle \mathbf{B}\left(\mathbf{x}\right)\cdot\mathbf{B}\left(\mathbf{x+\boldsymbol{\ell}}\right)\right\rangle \simeq B_{0}^{2}\left(\mathbf{x}\right)+\frac{1}{2}\left\langle \mathbf{B}_{0}\left(\mathbf{x}\right)\cdot\left(\mathbf{e}_{\ell}\cdot\nabla\right)^{2}\mathbf{B}_{0}\left(\mathbf{x}\right)\right\rangle \ell^{2}+\left\langle \mathbf{B}_{\mathrm{t}}\cdot\mathbf{B}_{\mathrm{t}}\left(\mathbf{\boldsymbol{\ell}}\right)\right\rangle ,\label{eq:num2}\end{equation}

\noindent when $\ell\ll d$.

Using the assumed homogeneity in the fields' strength (i.e., equations
{[}\ref{eq:homo}{]}) the denominator of equation (\ref{eq:cos(dphi)})
can be readily simplified to

\begin{eqnarray*}
\left[\left\langle B^{2}\left(\mathbf{x}\right)\right\rangle \left\langle B^{2}\left(\mathbf{x}+\boldsymbol{\ell}\right)\right\rangle \right]^{1/2} & = & \left\langle B^{2}\right\rangle \\
 & = & \left\langle B_{0}^{2}+B_{\mathrm{t}}^{2}+2\left(\mathbf{B}_{0}\cdot\mathbf{B}_{\mathrm{t}}\right)\right\rangle ,\end{eqnarray*}

\noindent which, with the fourth of equations (\ref{eq:averages}),
becomes

\begin{equation}
\left[\left\langle B^{2}\left(\mathbf{x}\right)\right\rangle \left\langle B^{2}\left(\mathbf{x}+\boldsymbol{\ell}\right)\right\rangle \right]^{1/2}=B_{0}^{2}+\left\langle B_{\mathrm{t}}^{2}\right\rangle .\label{eq:denom}\end{equation}

If we further assume isotropy over space (i.e., $\Delta\Phi_{\mathrm{3D}}\left(\boldsymbol{\ell}\right)=\Delta\Phi_{\mathrm{3D}}\left(\ell\right)$)
and insert equations (\ref{eq:num2}) and (\ref{eq:denom}) into equation
(\ref{eq:cos(dphi)}) we have

\begin{equation}
\left\langle \cos\left[\Delta\Phi_{\mathrm{3D}}\left(\ell\right)\right]\right\rangle \simeq1-\frac{\left\langle B_{\mathrm{t}}^{2}\right\rangle -\left\langle \mathbf{B}_{\mathrm{t}}\cdot\mathbf{B}_{\mathrm{t}}\left(\mathbf{\ell}\right)\right\rangle -\frac{1}{2}\left\langle \mathbf{B}_{0}\left(\mathbf{x}\right)\cdot\left(\mathbf{e}_{\ell}\cdot\nabla\right)^{2}\mathbf{B}_{0}\left(\mathbf{x}\right)\right\rangle \ell^{2}}{B_{0}^{2}+\left\langle B_{\mathrm{t}}^{2}\right\rangle },\label{eq:cos2}\end{equation}

\noindent when $\ell\ll d$. For cases where $\Delta\Phi_{\mathrm{3D}}\left(\ell\right)$
is small equation (\ref{eq:cos2}) simplifies to

\begin{equation}
\left\langle \Delta\Phi_{\mathrm{3D}}^{2}\left(\ell\right)\right\rangle \simeq\frac{2\left[\left\langle B_{\mathrm{t}}^{2}\right\rangle -\left\langle \mathbf{B}_{\mathrm{t}}\cdot\mathbf{B}_{\mathrm{t}}\left(\mathbf{\ell}\right)\right\rangle \right]}{B_{0}^{2}+\left\langle B_{\mathrm{t}}^{2}\right\rangle }-\frac{\left\langle \mathbf{B}_{0}\left(\mathbf{x}\right)\cdot\left(\mathbf{e}_{\ell}\cdot\nabla\right)^{2}\mathbf{B}_{0}\left(\mathbf{x}\right)\right\rangle }{B_{0}^{2}+\left\langle B_{\mathrm{t}}^{2}\right\rangle }\ell^{2},\label{eq:dphi2}\end{equation}

\noindent still when $\ell\ll d$.

Examining equation (\ref{eq:corr_Bt}) we recover the behavior of
the turbulent contribution to $\left\langle \Delta\Phi_{\mathrm{3D}}^{2}\left(\ell\right)\right\rangle $
(i.e., the first term on the right-hand side of equation {[}\ref{eq:dphi2}{]})
described in \S\ref{sec:Function} that goes from 0 when $\ell\rightarrow0$
to a constant, which we now define as $b_{\mathrm{3D}}^{2}$, when
$\ell>\delta$. The data sets analyzed in this paper are such that
$\ell>\delta$ in all cases. We therefore find that the dispersion
function is of the form

\begin{equation}
\left\langle \Delta\Phi_{\mathrm{3D}}^{2}\left(\ell\right)\right\rangle \simeq b_{\mathrm{3D}}^{2}+m_{\mathrm{3D}}^{2}\ell^{2},\label{eq:bml2}\end{equation}

\noindent with

\[
b_{\mathrm{3D}}^{2}=\frac{2\left\langle B_{\mathrm{t}}^{2}\right\rangle }{B_{0}^{2}+\left\langle B_{\mathrm{t}}^{2}\right\rangle }\]

\noindent when $\delta<\ell\ll d$. Once again, we identify $b_{\mathrm{3D}}$
with the constant contribution stemming from the turbulent field to
the total angular dispersion, while the larger scale contribution
due to variations in the large-scale field $\mathbf{B}_{0}$ is accounted
for by the presence of a term proportional to $\ell^{2}$ in equation
(\ref{eq:bml2}). 

\subsection{Analysis in Two Dimensions}

\label{sec:2D}The analysis presented above can still be used when
we limit ourselves to two dimensions. This is needed in order to enable
comparisons with polarization measurements, which only probe the plane-of-the-sky
component, $\mathbf{B}_{\Vert}$, of the magnetic field. Defining
$\mathbf{e}_{\bot}$ as the unit vector directed along the line-of-sight
we have for the total magnetic field

\begin{equation}
\mathbf{B}_{\Vert}=\mathbf{B}-\left(\mathbf{B}\cdot\mathbf{e}_{\bot}\right)\mathbf{e}_{\bot},\label{eq:Bpos}\end{equation}

\noindent and similar relations for $\mathbf{B}_{0}$ and $\mathbf{B}_{\mathrm{t}}$.

We need to evaluate, among others, the following autocorrelation

\begin{equation}
\left\langle \mathbf{B}_{\Vert}\left(\mathbf{x}\right)\cdot\mathbf{B}_{\Vert}\left(\mathbf{x+\boldsymbol{\ell}}\right)\right\rangle =\left\langle \mathbf{B}\left(\mathbf{x}\right)\cdot\mathbf{B}\left(\mathbf{x+\boldsymbol{\ell}}\right)\right\rangle -\left\langle \left[\mathbf{B}\left(\mathbf{x}\right)\cdot\mathbf{e}_{\bot}\right]\left[\mathbf{B}\left(\mathbf{x+\boldsymbol{\ell}}\right)\cdot\mathbf{e}_{\bot}\right]\right\rangle ,\label{eq:num_2d}\end{equation}

\noindent where the separation $\boldsymbol{\ell}$ is now limited
to the plane-of-the-sky. The last term on the right hand-side can
be transformed to

\begin{eqnarray}
\left\langle \left[\mathbf{B}\left(\mathbf{x}\right)\cdot\mathbf{e}_{\bot}\right]\left[\mathbf{B}\left(\mathbf{x+\boldsymbol{\ell}}\right)\cdot\mathbf{e}_{\bot}\right]\right\rangle  & = & \left\langle \left\{ \left[\mathbf{B}_{0}\left(\mathbf{x}\right)+\mathbf{B}_{\mathrm{t}}\left(\mathbf{x}\right)\right]\cdot\mathbf{e}_{\bot}\right\} \left\{ \left[\mathbf{B}_{0}\left(\mathbf{x}+\boldsymbol{\ell}\right)+\mathbf{B}_{\mathrm{t}}\left(\mathbf{x}+\boldsymbol{\ell}\right)\right]\cdot\mathbf{e}_{\bot}\right\} \right\rangle \nonumber \\
 & = & B_{0,\bot}^{2}+\left\langle \left[\mathbf{B}_{\mathrm{t}}\left(\mathbf{x}\right)\cdot\mathbf{e}_{\bot}\right]\left[\mathbf{B}_{\mathrm{t}}\left(\mathbf{x}+\boldsymbol{\ell}\right)\cdot\mathbf{e}_{\bot}\right]\right\rangle .\label{eq:corr_los}\end{eqnarray}

Using the same method that led to equation (\ref{eq:denom}) in the
three-dimensional case we also have that

\begin{equation}
\left[\left\langle B_{\Vert}^{2}\left(\mathbf{x}\right)\right\rangle \left\langle B_{\Vert}^{2}\left(\mathbf{x}+\boldsymbol{\ell}\right)\right\rangle \right]^{1/2}=B_{0,\Vert}^{2}+\left\langle B_{\mathrm{t},\Vert}^{2}\right\rangle .\label{eq:denom_2d}\end{equation}

We now introduce the function

\begin{equation}
\left\langle \cos\left[\Delta\Phi\left(\boldsymbol{\ell}\right)\right]\right\rangle \equiv\frac{\left\langle \mathbf{B}_{\Vert}\left(\mathbf{x}\right)\cdot\mathbf{B}_{\Vert}\left(\mathbf{x+\boldsymbol{\ell}}\right)\right\rangle }{\left[\left\langle B_{\Vert}^{2}\left(\mathbf{x}\right)\right\rangle \left\langle B_{\Vert}^{2}\left(\mathbf{x}+\boldsymbol{\ell}\right)\right\rangle \right]^{1/2}},\label{eq:cos_2d}\end{equation}

\noindent which upon inserting equations (\ref{eq:num2}), (\ref{eq:num_2d}),
(\ref{eq:corr_los}), and (\ref{eq:denom_2d}) with the condition
of space isotropy becomes

\[
\left\langle \cos\left[\Delta\Phi\left(\ell\right)\right]\right\rangle \simeq1-\frac{\left\langle B_{\mathrm{t},\Vert}^{2}\right\rangle -\left\langle \mathbf{B}_{\mathrm{t},\Vert}\cdot\mathbf{B}_{\mathrm{t},\Vert}\left(\mathbf{\ell}\right)\right\rangle -\frac{1}{2}\left\langle \mathbf{B}_{0}\left(\mathbf{x}\right)\cdot\left(\mathbf{e}_{\ell}\cdot\nabla\right)^{2}\mathbf{B}_{0}\left(\mathbf{x}\right)\right\rangle \ell^{2}}{B_{0,\Vert}^{2}+\left\langle B_{\mathrm{t},\Vert}^{2}\right\rangle },\]

\noindent when $\ell\ll d$ and where

\[
\left\langle \mathbf{B}_{\mathrm{t},\Vert}\cdot\mathbf{B}_{\mathrm{t},\Vert}\left(\mathbf{\ell}\right)\right\rangle =\left\langle \mathbf{B}_{\mathrm{t}}\cdot\mathbf{B}_{\mathrm{t}}\left(\mathbf{\ell}\right)\right\rangle -\left\langle \left[\mathbf{B}_{\mathrm{t}}\left(\mathbf{x}\right)\cdot\mathbf{e}_{\bot}\right]\left[\mathbf{B}_{\mathrm{t}}\left(\mathbf{x}+\boldsymbol{\ell}\right)\cdot\mathbf{e}_{\bot}\right]\right\rangle .\]

\noindent If we further consider $\Delta\Phi\left(\ell\right)$ to
be small, then we find 

\begin{equation}
\left\langle \Delta\Phi^{2}\left(\ell\right)\right\rangle \simeq\frac{2\left[\left\langle B_{\mathrm{t},\Vert}^{2}\right\rangle -\left\langle \mathbf{B}_{\mathrm{t},\Vert}\cdot\mathbf{B}_{\mathrm{t},\Vert}\left(\mathbf{\ell}\right)\right\rangle \right]}{B_{0,\Vert}^{2}+\left\langle B_{\mathrm{t},\Vert}^{2}\right\rangle }-\frac{\left\langle \mathbf{B}_{0}\left(\mathbf{x}\right)\cdot\left(\mathbf{e}_{\ell}\cdot\nabla\right)^{2}\mathbf{B}_{0}\left(\mathbf{x}\right)\right\rangle }{B_{0,\Vert}^{2}+\left\langle B_{\mathrm{t},\Vert}^{2}\right\rangle }\ell^{2}\label{eq:dphi2_2d}\end{equation}

\noindent still when $\ell\ll d$ and the displacement $\ell$ is
limited to the plane-of-the-sky. 

For our data sets we have the further simplification that $\delta<\ell\ll d$
and the dispersion function, equation (\ref{eq:dphi2_2d}), has then
a form similar to its three-dimensional counterpart with

\begin{equation}
\left\langle \Delta\Phi^{2}\left(\ell\right)\right\rangle \simeq b^{2}+m^{2}\ell^{2},\label{eq:bmpos}\end{equation}

\noindent where 

\begin{equation}
b^{2}=\frac{2\left\langle B_{\mathrm{t},\Vert}^{2}\right\rangle }{B_{0,\Vert}^{2}+\left\langle B_{\mathrm{t},\Vert}^{2}\right\rangle }\label{eq:bpos}\end{equation}

\noindent is the quantity we evaluate through polarization measurements.
Equation (\ref{eq:bpos}) can be transformed to yield the ratio of
the turbulent to large-scale magnetic field strength through

\begin{equation}
\frac{\left\langle B_{\mathrm{t},\Vert}^{2}\right\rangle ^{1/2}}{B_{0,\Vert}}=\frac{b}{\sqrt{2-b^{2}}}.\label{eq:ratio}\end{equation}

\section{Data Analysis}\label{sec:Data}

Data from the Hertz polarimeter on the clouds studied here have been
previously published by \citet{Houde2004b} for OMC-1, \citet{Houde2002}
for M17, and \citet{Kirby2009} for DR21(Main). Details on the instrument
as well as data acquisition and reduction can be found in \citet{Dowell1998}
and \citet{Kirby2005}, respectively. The analysis presented here
is performed on a complete re-reduction of the raw Hertz data using
the method of \citet{Kirby2005} and \citet{Dotson2009}. The data
may differ slightly from that published in the references above. For
our purposes we only include data which satisfy the $P>3\sigma_{P}$
criterion, where $P$ is the polarization fraction and $\sigma_{P}$
its uncertainty.

In each of the three objects the angle differences between each and
every pair of data points are calculated as

\begin{equation}
\Delta\Phi_{ij}=\Phi_{i}-\Phi_{j},\label{eq:diff}\end{equation}
and the corresponding distance between each point

\begin{equation}
\ell_{ij}\equiv\vert\mathbf{x}_{i}-\mathbf{x}_{j}\vert.\label{eq:distance}\end{equation}

\noindent Note that $\ell_{ij}=\ell_{ji}$ so that a map with $N$
data points contains only $N(N-1)/2$ distinct differences. Also note
that $\left|\Delta\Phi_{ij}\right|$ is constrained to be in the range
$[0,90]$ degrees.

These data are divided into separate distance bins with sizes corresponding
to integer multiples of a single Hertz pixel-to-pixel separation ($17\farcs8$);
the first bin covers $\ell_{1}\leq\ell_{ij}<\ell_{2}$ (where $\ell_{k}$
corresponds to $k$ pixels). Within each bin $k$ we calculate the
dispersion as the root-mean-square of the angle difference

\begin{equation}
\left\langle \Delta\Phi_{ij}^{2}\right\rangle _{k}^{1/2}\,\,\,\,\,\mathrm{for\, all\,\,\,\,}\ell_{k}\leq\ell_{ij}<\ell_{k+1}.\label{eq:rms}\end{equation}

The dispersion is corrected for measurement uncertainty within each
bin according to equation (\ref{eq:disp_m}). The uncertainty on each
$\Delta\Phi_{ij}$ follows from simply propagating the measurement
uncertainties on both $\Phi_{i}$ and $\Phi_{j}$ through equation
(\ref{eq:diff}). The root-mean-square measurement uncertainties within
each bin $k$ are then given by

\[
\sigma_{\mathrm{M},k}=\left\langle \sigma^{2}(\Delta\Phi_{ij})\right\rangle _{k}^{1/2}\,\,\,\,\,\mathrm{for\, all\,\,\,\,}\ell_{k}\leq\ell_{ij}<\ell_{k+1}.\]

\begin{figure}
\epsscale{0.7}
\plotone{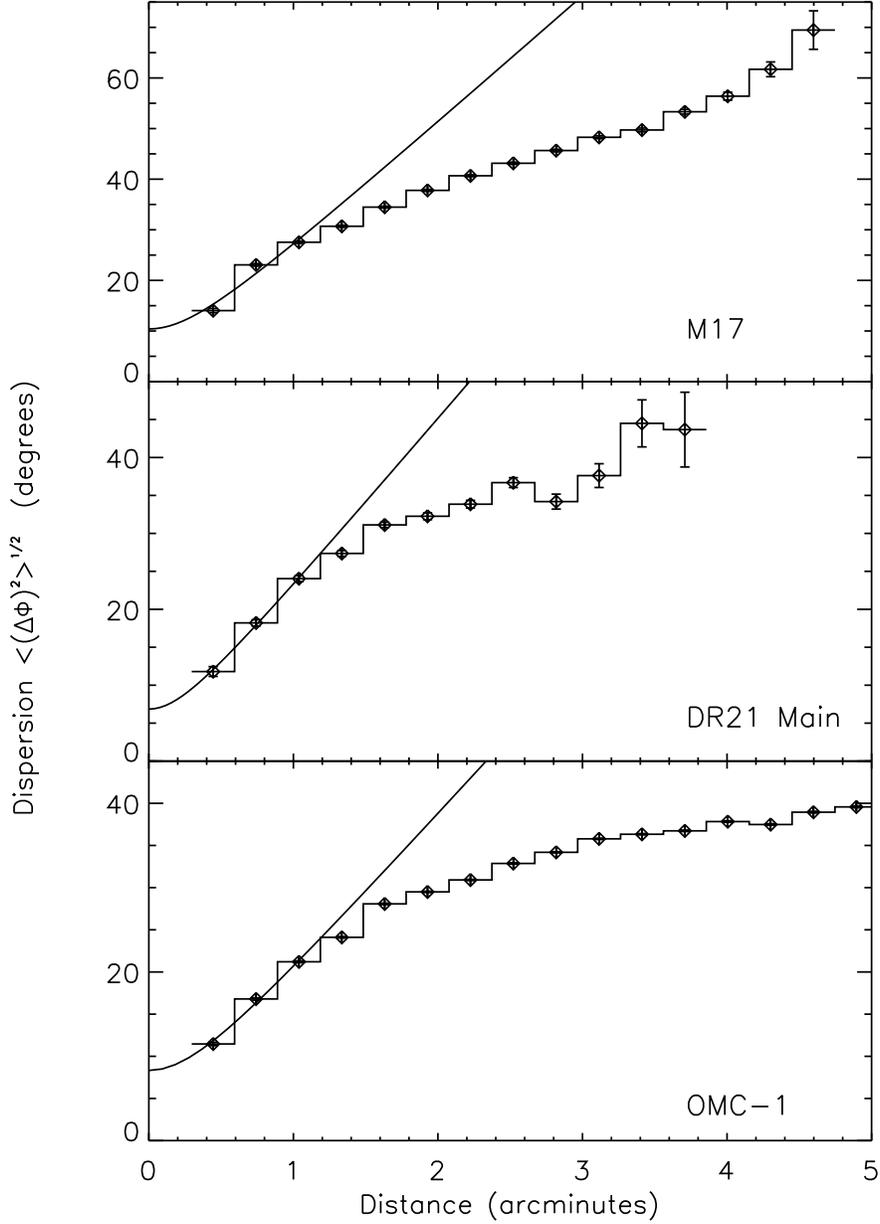}

\caption{\label{fig:data}Angular dispersion function, $\left\langle \Delta\Phi^{2}\left(\ell\right)\right\rangle ^{1/2}$,
for M17, DR21(Main), and OMC-1. The turbulent contribution to the
total angular dispersion is determined by the zero intercept of the
fit to the data at $\ell=0$. The measurement uncertainties were removed
prior to operating the fits to the corresponding data sets. The results
are given in Table \ref{tab:Results}. }

\end{figure}

The corrected dispersions are those plotted for the different bins
in Figure \ref{fig:data}. The error bars in Figure \ref{fig:data}
are determined by propagating the measurement uncertainties for $\Phi_{i}$
and $\Phi_{j}$ through equations (\ref{eq:diff}) and (\ref{eq:rms}),
most of these are too small to be seen in the figure, especially at
the smallest displacements. 

For each object, the data are fitted to a linear model of the corrected
square dispersion with respect to the square of the distance according
to equation (\ref{eq:disp}). In the discrete notation introduced
in this section, the model is given by

\[
\left\langle \Delta\Phi_{ij}^{2}\right\rangle _{k}-\sigma_{\mathrm{M},k}^{2}=b^{2}+m^{2}\ell_{k}^{2}.\]

In order to ensure we are in the linear regime, the fits are limited
to only the smallest three distance bins. Taking into account the
uncertainties on the $\left\langle \Delta\Phi_{ij}^{2}\right\rangle _{k}$,
the least-squares solutions for the parameter $b$ are given in Table
\ref{tab:Results}. 

\end{document}